\documentstyle[12pt,epsf]{article}

\renewcommand{\footnote}[1]{%
\def\thefootnote{\arabic{footnote})}
\footnotemark%
\footnotetext{#1}}
\newcommand{\mod}{{\rm mod}}
\newcommand{\sgn}{{\rm sgn}}
\newcommand{\arctg}{{\rm arctg}}
\newcommand{\tg}{{\rm tg}}
\newcommand{\ctg}{{\rm ctg}}
\newcommand{\io}{[\hspace{-1pt}[}
\newcommand{\ic}{]\hspace{-1pt}]}
\newcommand{\fo}{\{\!\mid}
\newcommand{\fc}{\mid\!\}}
\newcommand{\sint}{{\setbox1=\hbox{$\displaystyle\sum$}\displaystyle\sum\kern-.95\wd1\int\limits_E\,}}
\newcommand{\vV}{{\bf V}}
\newcommand{\vx}{{\bf x}}
\newcommand{\valpha}{\mbox{\boldmath $\alpha$}}
\newcommand{\vgamma}{\mbox{\boldmath $\gamma$}}

\title{Quantum Numbers of the $\Theta$ Vacuum in (2+1)-Dimensional 
  Spinor Electrodynamics: Charge and Magnetic Flux.{\footnote{Published 
  in \em Physics of Atomic Nuclei, Vol. 60, N 9, 1997, pp. 1497-1503.}}}

\author{Yu. A. Sitenko{\footnote{e-mail: yusitenko@gluk.apc.org}},
  \qquad D. G. Rakityansky{\footnote{e-mail: radamir@ap3.gluk.apc.org}}\\
  Bogolyubov Institute for Theoretical Physics,\\
  National Academy of Sciences of Ukraine,\\
  252143 Kiev, Ukraine}

\begin{document}
\maketitle

\begin{abstract}
A singular configuration of an external static vector field in the form of a 
magnetic string polarizes the vacuum of a second-quantized theory on the 
plane orthogonal to the string axis. The most general boundary conditions at 
the punctured singular point that are compatible with the self-adjointness of 
two-dimensional Dirac Hamiltonian are considered. The dependences of the 
induced vacuum quantum numbers on the parameter of the self-adjoint 
extension, on the string flux, and on the choice of irreducible 
reoresentation of the matrices in (2+1)-dimensional spacetime are disscussed.
\end{abstract}

\section*{}
{\bf 1.}
It is well known that, in the fermion vacuum, a singular magnetic monopole 
induces the charge \cite{Wit, Gro, Yam}
\begin{equation}\label{1}
Q^{(I)}=-\frac{1}{\pi}{\rm arctg}\left({\rm tg}{\frac{\Theta}{2}}\right),
\end{equation}
where $\Theta$ is the self-adjoint-extension parameter that determines the 
boundary conditions at the punctured point corresponding to the monopole 
position. As a result, the monopole actually becomes a dion that violates 
{\em CP} symmetry and the condition of Dirac quantization.

In this study, we consider quantum numbers induced in the fermion vacuum by a 
singular static magnetic string. In relation to the elimination of a point, 
the elimination of a line leads to more substantial changes in the topology 
of space (the first homotopic group becomes nontrivial). Therefore, the 
properties of the $\Theta$ vacuum are much richer in this case than in the 
case specified by (\ref{1}). We will restrict our consideration to the plane 
orthogonal to the plane axis and study (2+1)-dimensional spinor 
electrodynamics on the plane with the punctured point corresponding to the 
string position. It can be shown \cite{Sit96,Sit96rus} that, in this case, 
charge and magnetic flux are induced in the vacuum. These quantities deppend 
both on the parameter of the self-adjoint extension and on the magnetic flux 
of the string. In this study, we perform a more detailed analysis of these 
dependences.

\section*{}
{\bf 2.}  
Let us write the time-independent Dirac equation in an extternal vector field
$\vV(\vx)$ in tthe form
\begin{equation}\label{2}
\left\{-i\valpha\left[\frac{\partial}{\partial\vx}-i\vV(\vx)\right]
+\beta m\right\}\psi(\vx)=E\psi(\vx),
\end{equation}
where
\begin{equation}\label{3}
\valpha=\gamma^0\vgamma,\qquad\beta=\gamma^0,
\end{equation}
and $\vgamma$ and  $\gamma^0$ are the Dirac $\gamma$ matrices.
In the (2+1)-dimensional spacetime $(\vx,t)=(x^1,x^2,t)$,
the Clifford allgebra does nott have a faithful irreducible representation.
Instead, it has two nonequivalent irreduceble representations? which differ 
from one another in the following way:
\begin{equation}\label{4}
i\gamma^0\gamma^1\gamma^2=s,\qquad s=\pm 1.
\end{equation}
Choosing the matrix $\gamma^0$ in the diagonal form 
\begin{equation}\label{5}
\gamma^0=\sigma_3,
\end{equation}
we obtain
\begin{equation}\label{6}
\gamma^1=e^{\frac{i}{2}\sigma_3\chi_s}i\sigma_1
e^{-\frac{i}{2}\sigma_3\chi_s},
\qquad
\gamma^2=e^{\frac{i}{2}\sigma_3\chi_s}is\sigma_2
e^{-\frac{i}{2}\sigma_3\chi_s},
\end{equation}
where $\sigma_1$, $\sigma_2$ and $\sigma_3$ are the Pauli matrices,
and $\chi_1$ and $\chi_{-1}$ are parameters that are varied in the interval 
$0\leq\chi_s<2\pi$ to go over to equivalent representations.

We choose the configuration of the external field
$\vV(\vx)=\left(V_1(\vx),V_2(\vx)\right)$
in a form of a singular magnetic vortex
\begin{equation}\label{7}
\vx\times\vV(\vx)=\Phi^{(0)},\qquad
B(\vx)\equiv\frac{\partial}{\partial\vx}\times\vV(\vx)=2\pi\Phi^{(0)}\delta(\vx),
\end{equation}
where $\Phi^{(0)}$ is the total flux (in $2\pi$ units) of the vortex --- that is,
of the string that intersects the plane at the point $\vx=0$.
The wave function on the plane with the punctured singular point
$\vx=0$ obeys the most general condition (see. \cite{Sit96} for more details)
\begin{equation}\label{8}
\psi(r,\phi+2\pi)=e^{i2\pi\Upsilon}\psi(r,\phi),
\end{equation}
where $r=\mid \vx\mid$ and $\phi=\arctg(x^2/x^1)$ are polar coordinates, and
$\Upsilon$ is a continuos real parameter.

A solution that satisfies the Dirac equation (\ref{2}) in the field of the 
singular string (\ref{7}) and condition (\ref{8}) can be represented
as 
\begin{equation}\label{9}
\psi(\vx)=\sum_{n=-\infty}^\infty
\left(
\begin{array}{l}
f_n(r)e^{i(n+\Upsilon)\phi}\\
g_n(r)e^{i(n+\Upsilon+s)\phi}
\end{array}
\right), 
\end{equation}
where radial functions $f_n(r)$ and $g_n(r)$ obey the system of equations
\begin{equation}\label{10}
\begin{array}{l}
e^{-i\chi_s}[-\partial_r+s(n-\Phi^{(0)}+\Upsilon)r^{-1}]f_n(r)=(E+m)g_n(r),\\
\\
e^{i\chi_s}[\partial_r+s(n-\Phi^{(0)}+\Upsilon+s)r^{-1}]g_n(r)=(E-m)f_n(r).
\end{array}
\end{equation}
For integer values of $\Phi^{(0)}-\Upsilon$, the condition of square 
integrability makes it possible to construct solutions for the top and bottom
components ($f_n(r)$ and  $g_n(r)$, respectively) of the spinor wave function
in such way that they are regular at the point $r=0$.
When $\Phi^{(0)}-\Upsilon$ is fractional, the same situation occures only for
$n\neq n_0$, where
\begin{equation}\label{11}
n_0=\io\Phi^{(0)}-\Upsilon\ic+\frac{1}{2}-\frac{1}{2}s;
\end{equation}
here $\io u\ic$ is the integral part of $u$ (the largest integer that is less 
than or equal to $u$). For $n=n_0$, each of linear independent solutions 
satisfies the requirement of square integrability. If one solution is chosen to 
have a regular top and a singular bottom component, the other has a regular 
bottom and a singular top component. In other words, the partial Dirac 
Hamiltonian, corresponding to $n\neq n_0$ represents the family of self-adjoint 
extensions parametrized by one continuous real variable ($\Theta$).
It follows that, instead of satisfying the regularity condition at the point
$r=0$, the radial functions for $n=n_0$, obey the condition \cite{Ger} 
\begin{equation}\label{12}
\lim_{r\rightarrow 0}(\mid m\mid r)^F\cos\left(\frac{\Theta}{2}+\frac{\pi}{4}
\right)f_{n_0}(r)=-e^{i\chi_s}
\lim_{r\rightarrow 0}(\mid m\mid r)^{1-F}\sin\left(\frac{\Theta}{2}+\frac{\pi}{4}
\right)g_{n_0}(r),
\end{equation}
where
\begin{equation}
F=\frac{1}{2}+s\left(\fo\Phi^{(0)}-\Upsilon\fc-\frac{1}{2}\right);
\end{equation}
here $\fo u\fc=u-\io u\ic$ is the fractional part of $u$, $0\leq\fo u\fc<1$.
It is wortth noting, in (2+1)-dimensional space-time, the mass parameter
$m$ in the Dirac equation (\ref{2}) can take botth positive and negative values.
We emphasize once again that relation (\ref{12}) holds only for $0<F<1$, 
because? in tthe case of $F=0$ (integral values of $\Phi^{(0)}-\Upsilon$),
the radial functions are regular both for $n=n_0$ and for $n\neq n_0$.

Suppose that $0<F<1$.  When the condition
\begin{equation}\label{-infinity}
-\infty<\sgn(m)\tg\left(\frac{\Theta}{2}+\frac{\pi}{4}\right)<0,
\end{equation}
is satisfied, the spectrum of states involves not only a continuum ($|E|>|m|$)
but also a bound sttate ($|E|<|m|$) whose energy is determined by the equation
\cite{Sit96rus}
\begin{equation}\label{BS}
\frac{(1+m^{-1}E)^{1-F}}{(1-m^{-1}E)^F}=-\sgn(m)A;
\end{equation}
where
\begin{equation}\label{A}
A=2^{1-2F}\frac{\Gamma(1-F)}{\Gamma(F)}\tg\left(\frac{\Theta}{2}+
\frac{\pi}{4}\right),
\end{equation}
and
\[
\sgn(u)=\left\{
\begin{array}{rl}
 1,&u<0\\
-1,&u>0
\end{array}
\right..
\]
In equation (\ref{A}), $\Gamma(u)$ is the Euler gamma-function.

It can be seenfrom (\ref{BS}) that the energy of the bound state vanishes 
($E=0$) when
\begin{equation}\label{EO}
\sgn(m)A=-1.
\end{equation}

\section*{}
{\bf 3.} 
The expressions for the density of the vacuum charge and for the angular component
of the vacuum current are, respectively,
\begin{equation}\label{14}
\rho(r)=-\frac{1}{2}\sint\sum_{n=-\infty}^{\infty}\sgn(E)[f_n^*(r)f_n(r)
+g_n^*(r)g_n(r)]
\end{equation}
and
\begin{equation}\label{15}
j^\phi(r)=-e^{i\chi_s}\sint\sum_{n=-\infty}^{\infty}\sgn(E)f_n^*(r)g_n(r),
\end{equation}
where the symbol $\sint$ denotes summation over discrete values of the energy 
$E$ and integration (with a srtain measure) over its continuous values.
The radial component of the vacuum current vanishes identically by virtue 
of the condition
\begin{equation}\label{16}
e^{i\chi_s}f_n^*(r)g_n(r)=e^{-i\chi_s}g_n^*(r)f_n(r).
\end{equation}
Integrating (\ref{14}) over the entire two-dimensional space, we obtaine the 
total vacuum charge
\begin{equation}\label{17}
Q^{(I)}=2\pi\int_0^\infty dr\, r\rho(r).
\end{equation}
A global charachteristic associated with the vacuum current is the total 
magnetic flux of the vacuum. 
In $2\pi$ units, this magnetic flux is given by 
\begin{equation}\label{18}
\Phi^{(I)}=\frac{e^2}{2}\int_0^\infty dr\, r^2j^\phi(r),
\end{equation}
where $e$ is the coupling constant having dimensions of  $\sqrt{\mid m\mid}$
in (2+1)-dim\-en\-sional space-time. In deriving relation (\ref{18}),
we assumed that the vacuum magnetic field is related to the vacuum current 
by the Maxwell equation
\begin{equation}\label{19}
\partial_rB_{(I)}(r)=-e^2j^\phi(r),
\end{equation}
and that the vacuum current decreases sufficiently fast (exponentially --- see
\cite{Sit96rus}) for $r\rightarrow\infty$.

By using explicit expressions for the solutions to the system of equations
(\ref{10}), we can find the functional dependence of the vacuum quantum numbers 
(\ref{17}) and (\ref{18}) on the parameter of self-adjoint extension and on the 
magnetic flux of the string. It is obvious that, for a fixed value of $\Theta$,
the vacuum quantum numbers depend periodically on $\Phi^{(0)}-\Upsilon$, the 
period being equal to unity. For the integral values of $\Phi^{(0)}-\Upsilon$, 
the radial functions $f_n(r)$ and $g_n(r)$ are regular at the point $r=0$,
and this case is indistinguishable from that of the trivial vacuum 
$\Phi^{(0)}=\Upsilon=0$. As a result, we obtain
\begin{equation}\label{20}
Q^{(I)}=\Phi^{(I)}=0,\qquad F=0.
\end{equation}
It was noted above that, for nonintegral values of $\Phi^{(0)}-\Upsilon$, 
the radial functions are regular for $n\neq n_0$ and satisfy condition (\ref{12})
at $n=n_0$. Thus, we have \cite{Sit96,Sit96rus}
\begin{equation}\label{21}
Q^{(I)}=\left\{
\begin{array}{ll}
-\frac{1}{2}\sgn(m)F    ,&\Theta=\frac{\pi}{2}\mod 2\pi\\&\\
 \frac{1}{2}\sgn(m)(1-F),&\Theta=\left(-\frac{\pi}{2}\right)\mod 2\pi\\&\\
-\sgn(m)\left[\frac{1}{2}\left(F-\frac{1}{2}\right)+S_1(F,\Theta)\right]-
S_2(F,\Theta),&\Theta\neq\left(\pm\frac{\pi}{2}\right)\mod 2\pi
\end{array}\right.
\end{equation}
\begin{equation}\label{22}
\Phi^{(I)}=-\frac{e^2F(1-F)}{2\pi\mid m\mid}\left[\frac{1}{6}\left(F-
\frac{1}{2}\right)+S_1(F,\Theta)\right],
\end{equation}
where
\begin{equation}
\label{23}
S_1(F,\Theta)=\frac{1}{4\pi}\int_1^\infty\frac{dv}{v\sqrt{v-1}}
\frac{Av^F-A^{-1}v^{1-F}}{Av^{F}+2\sgn(m)+A^{-1}v^{1-F}},
\end{equation}
\begin{equation}\label{24}
S_2(F,\Theta)=\frac{F-\frac{1}{2}}{\pi}\int_1^\infty\frac{dv}{v}
\frac{\sqrt{v-1}}{Av^{F}+2\sgn(m)+A^{-1}v^{1-F}},
\end{equation}
and the quality $A$ is determined by (\ref{A}).

As might have been expected, the vacuum quantum numbers (\ref{21}) and
(\ref{22}) remain unchanged upon going over to an equivalent representation 
(they do not depend on the parameter $\chi_s$) and, in general, change upon 
going over to a nonequivalent representation (the substitution $s\rightarrow -s$ 
is equivalent to the substitution $F\rightarrow 1-F$). In the following, we 
treat the variables $\Theta$ and $F$as independent ones.
The substitution $m\rightarrow -m$ in tthe expressions for $Q^{(I)}$
and $\sgn(m)\Phi^{(I)}$ is then equivalent to the simultaneous substittutions 
$\Theta\rightarrow \Theta+\pi$ and $F\rightarrow 1-F$.

The function $S_2(F,\Theta)$ in (\ref{24}) can be represented in the form
\begin{equation}\label{26}
S_2(F,\Theta)=\left\{
\begin{array}{ll}\displaystyle
-\frac{1}{\pi}\int_0^1\frac{dw\sqrt{1-w^{(\frac{1}{2}-F)^{-1}}}}
{Aw^2+2\sgn(m)w^{(1-F)(\frac{1}{2}-F)^{-1}}+A^{-1}},&0<F<\frac{1}{2}\\&\\
0,&F=\frac{1}{2}\\&\\ \displaystyle
\frac{1}{\pi}\int_0^1\frac{dw\sqrt{1-w^{(F-\frac{1}{2})^{-1}}}}
{A^{-1}w^2+2\sgn(m)w^{F(F-\frac{1}{2})^{-1}}+A},&\frac{1}{2}<F<1
\end{array}\right.
\end{equation}
The following relations also hold:
\begin{eqnarray}
\frac{\partial}{\partial\Theta}[\sgn(m)S_1(F,\Theta)+S_2(F,\Theta)]&=&
\lim_{v\rightarrow\infty}\frac{1}{\pi\cos\Theta}
\frac{\sqrt{v-1}}{Av^F+2\sgn(m)+A^{-1}v^{1-F}}
=\nonumber\\\label{27}
&=&\left\{
\begin{array}{ll}
0,&F\neq\frac{1}{2}\\&\\
\frac{1}{2\pi},&F=\frac{1}{2}
\end{array}
\right.,
\end{eqnarray}
\begin{eqnarray}
\frac{\partial}{\partial F}[\sgn(m)S_1(F,\Theta)+S_2(F,\Theta)]&=&
\lim_{v\rightarrow\infty}\frac{1}{\pi}
\frac{\sqrt{v-1}\left(\pi\ctg F\pi+\ln\frac{v}{4}\right)}{Av^F+2\sgn(m)+A^{-1}v^{1-F}}
=\nonumber\\\label{d/dF}
&=&\left\{
\begin{array}{ll}
0,&F\neq\frac{1}{2}\\&\\
0,&\Theta=\left(\pm\frac{\pi}{2}\right)\mod2\pi,F=\frac{1}{2}\\&\\
\infty,&\Theta\neq\left(\pm\frac{\pi}{2}\right)\mod2\pi,F=\frac{1}{2}
\end{array}.
\right.
\end{eqnarray}

In the case of $F=\frac{1}{2}$ we can obttain (compare with (\ref{1}))
\begin{equation}\label{28}
Q^{(I)}=-\frac{1}{2\pi}\arctg\left\{\tg\left[
\frac{\Theta}{2}+\frac{\pi}{4}(1-\sgn(m))
\right]\right\},
\end{equation}
\begin{equation}\label{29}
\Phi^{(I)}=-\frac{e^2}{8\pi^2m}\arctg\left\{\tg\left[
\frac{\Theta}{2}+\frac{\pi}{4}(1-\sgn(m))
\right]\right\}.
\end{equation}

\section*{}
{\bf 4.}
The vacuum magnetic flux considered as a function of $\Theta$ has a discontinuity 
at $\Theta=\Theta_0$, where
\begin{equation}\label{30}
\Theta_0=\left\{\frac{3\pi}{2}-2\sgn(m)\arctg\left[2^{2F-1}
\frac{\Gamma(F)}{\Gamma(1-F)}\right]\right\}\mod 2\pi;
\end{equation}

At this value of $\Theta$, the energy of the bound state vanishes  
(see (\ref{EO})).  The vacuum charge, as a function of $\Theta$, has 
discontinuities not only at $\Theta=\Theta_0$ but also at $\Theta=\Theta_-$ 
for $0<F<\frac{1}{2}$ ¨and at $\Theta=\Theta_+$ for $\frac{1}{2}<F<1$, where
\begin{equation}\label{31}
\Theta_{\pm}=\left(\mp\frac{\pi}{2}\right)\mod 2\pi;
\end{equation}
By virtue of relation (\ref{27}), the vacuum charge is a constant for 
$F\neq\frac{1}{2}$. It should be emphasized that the vacuum quantum numbers 
are indeterminate at $\Theta=\Theta_0$ given by (\ref{30}).

For $\Theta$ values from the region determined by (\ref{-infinity}), the vacuum 
quantum numbers as functions of $F$ have discontinuities at the $F$ value 
satisfying relation (\ref{EO}). The values of the vacuum quantum numbers are 
indeterminate at this point discontinuity. In addition, tthe vacuum charge as 
a function of $F$ has a discontinuity at $F=\frac{1}{2}$, provided that 
$\Theta\neq\left(\pm \frac{\pi}{2}\right)\mod2\pi$:
\begin{equation}\label{32}
Q^{(I)}\mid_{F\nearrow\frac{1}{2}}=
Q^{(I)}\mid_{F=\frac{1}{2}}+\frac{1}{\pi}\arctg\left[\tg\left(
\frac{\Theta}{2}+\frac{\pi}{4}\right)\right],
\end{equation}
\begin{equation}\label{33}
Q^{(I)}\mid_{F\searrow\frac{1}{2}}=
Q^{(I)}\mid_{F=\frac{1}{2}}+\frac{1}{\pi}\arctg\left[\tg\left(
\frac{\Theta}{2}+\frac{\pi}{4}\right)\right]
-\frac{1}{2}\sgn\left[\tg\left(\frac{\Theta}{2}+\frac{\pi}{4}\right)\right];
\end{equation}
When $\Theta=\{\frac{\pi}{2}[1+\sgn(m)]\}\mod2\pi$, the two discontinuities 
coinside, and the value of the vacuum charge is indeterminate at this point.

Taking into account (\ref{d/dF}), (\ref{32}) and (\ref{33}), we find that,
for $F\neq\frac{1}{2}$, the vacuum charge can be represented as
\begin{equation}\label{F1/2}
Q^{(I)}=\left\{
\begin{array}{ll}
\hphantom{-}\frac{1}{2}\sgn(m)(1-F),&-1<\sgn(m)A<\infty\\&\\
-\frac{1}{2}\sgn(m)(1+F),&-\infty<\sgn(m)A<-1\\&\\
-\frac{1}{2}\sgn(m)F,&A^{-1}=0
\end{array}\right\},\quad 0<F<\frac{1}{2};
\end{equation}
\begin{equation}\label{F1}
Q^{(I)}=\left\{
\begin{array}{ll}
-\frac{1}{2}\sgn(m)F,&-1<\sgn(m)A^{-1}<\infty\\&\\
\hphantom{-}\frac{1}{2}\sgn(m)(2-F),&-\infty<\sgn(m)A^{-1}<-1\\&\\
\hphantom{-}\frac{1}{2}\sgn(m)(1-F),&A=0
\end{array}\right\},\quad \frac{1}{2}<F<1.
\end{equation}

The vacuum quantum numbers are displayed in Figs.1 and 2 for $m>0$ 
and in Figs. 3 and 4 for $m<0$. As can be seen from (\ref{-infinity}), a bound
state exists for $\frac{\pi}{2}<\Theta<\frac{3\pi}{2}$ in the former case 
($\mid E\mid<m$) and for $-\frac{\pi}{2}<\Theta<\frac{\pi}{2}$in the latter 
case ($\mid E\mid<-m$).

\section*{}
{\bf 5.}
In this study, we consider the vacuum quanttum numbers for the most general 
boundary conditions at the point $r=0$. These conditions can violate 
C symmetry. Under charge conjugation, we have $\Phi^{(0)}\rightarrow -\Phi^{(0)}$
and $\Upsilon\rightarrow-\Upsilon$; this is equivalent to the substitution
$F\rightarrow 1-F$. In the case of boundary conditions conserving C symmetry,
we have $Q^{(I)}\rightarrow -Q^{(I)}$ and $\Phi^{(I)}\rightarrow -\Phi^{(I)}$.
Several examples of boundary conditions conserving C symmetry are represented in 
\cite{Sit96}. For one boundary condition of this type, the condition of weaker 
singularity of wave function at the point $r=0$ (that is, the condition 
requiring that the divergence for $r\rightarrow 0$ not be stronger than 
$r^{-p}$, where $p\leq \frac{1}{2}$), the vacuum charge was calculated in 
\cite{SitR}.\footnote{For the condition of weaker singularity, the vacuum 
angular momentum was calculated in \cite{Sit90}.} 
It is clear from our consideration (see Figs. 1 and 3) that any boundary 
condition corresponding to $-\frac{\pi}{2}\sgn(m)<\Theta<\pi-
\frac{\pi}{2}\sgn(m)$ for $F\neq\frac{1}{2}$ ($0<\sgn(m)A<\infty$) and
$\Theta=\frac{\pi}{2}[1-\sgn(m)]$ for $F=\frac{1}{2}$ ($\sgn(m)A=1$) conservs
C symmetry for the vacuum charge. In this case, however, C symmetry is violated
for the vacuum magnetic flux (see Figs. 2 and 4). The boundary conditions that 
conserve C symmetry and periodicity in the magnetic flux of the string both for 
the vacuum charge and for the magnetic flux are given by
\begin{equation}
\left.
\begin{array}{lc}
\Theta =\Theta_C\mod 2\pi,&0<F<\frac{1}{2}\\&\\
\Theta=\left\{\frac{\pi}{2}[1-\sgn(m)]\right\}\mod 2\pi,& F=\frac{1}{2}\\&\\
\Theta=(-\Theta_C)\mod 2\pi,& \frac{1}{2}<F<1
\end{array}
\right\},
-\pi<\Theta_C\leq\pi.
\end{equation}

It is clear that, by choosing a different $\Theta$ values for different values 
of the magnetic flux of the string, we can specify boundary conditions 
violating the periodicity of the vacuum numbers in $\Phi^{(0)}$. 
Accordingly, there exist a number of boundary conditions that conserve C symmetry 
and which violate periodicity in  $\Phi^{(0)}$ (an example of this type is 
considered in \cite{Sit96}).  

Returning to boundary conditions that are periodic in $\Phi^{(0)}$,
we emphasize that, by taking an appropriate value of $\Theta$ from the interval 
$-\frac{\pi}{2}\sgn(m)<\Theta<\pi-\frac{\pi}{2}\sgn(m)$ for each value of $F$
from the interval $0<F<1$, we can annihilate the vacuum magnetic flux for all 
values of  $\Phi^{(0)}$ (see Figs. 2 and 4).  However, it is impossible to 
achieve this for the vacuum charge (see Figs. 1 and 3). Hence, there is no 
physically acceptable booundary condition (that is, a boundary condition 
compatible with the self-adjointness of the Hamiltonian) that ensures the
vanishing of the vacuum quantum numbers for all values of the magnetic flux 
of the string.\footnote{The opposite statement in \cite{Pol} is erroneous.}

In conclusion, we note that the region of the vacuum quantum numbers is restricted 
by the conditions
\begin{equation}
-\frac{3}{4}<Q^{(I)}<\frac{3}{4},\qquad-\frac{4}{5}<\frac{2\pi|m|}{e^2F(1-F)}
\Phi^{(I)}<\frac{4}{5};
\end{equation}
I this case, we can therefore speak about the fractional charge in the primary 
sense of this notion \cite{Jack76,Gold}.
That the vacuum charge grows indefinitely with increasing magnetic flux of the 
string, as was stated in \cite{Par,Fle}, is incorrect.

\section*{}
This work was supported in part by the State Commettee for Science, Technologies,
and Industrial Policy of Ukraine and by the program Few-Body Physics Network 
(INTAS-93-337).

\begin{center}
\epsfxsize=16cm
\epsffile{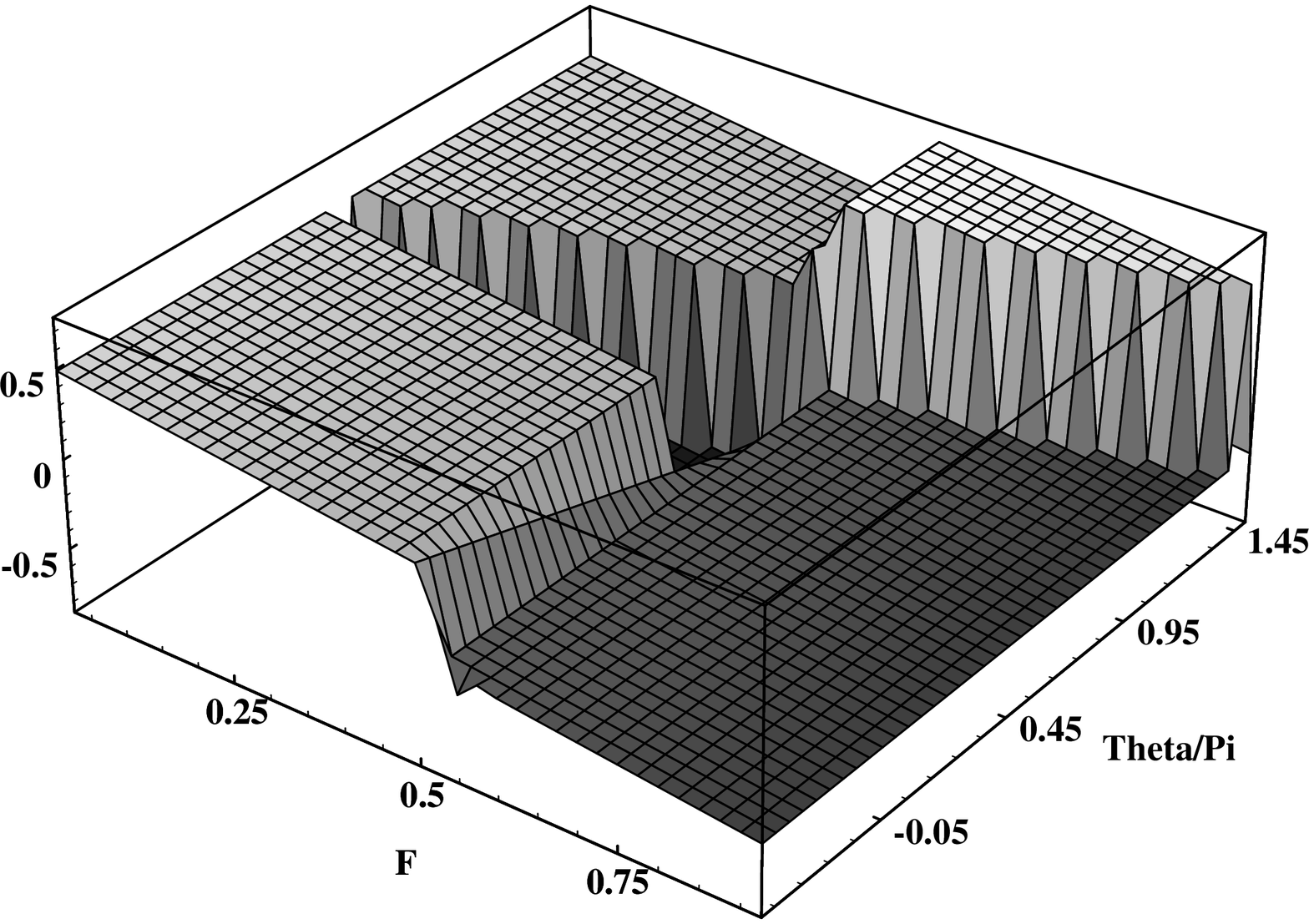}
\leavevmode
\end{center}
\vfill
\noindent Fig. 1.\qquad $Q^{(I)}$ in the region $-0.5<\Theta\pi^{-1}<1.5$, $0<F<1$
($m>0$).
\newpage
\begin{center}
\thispagestyle{empty}
\epsfxsize=16cm
\epsffile{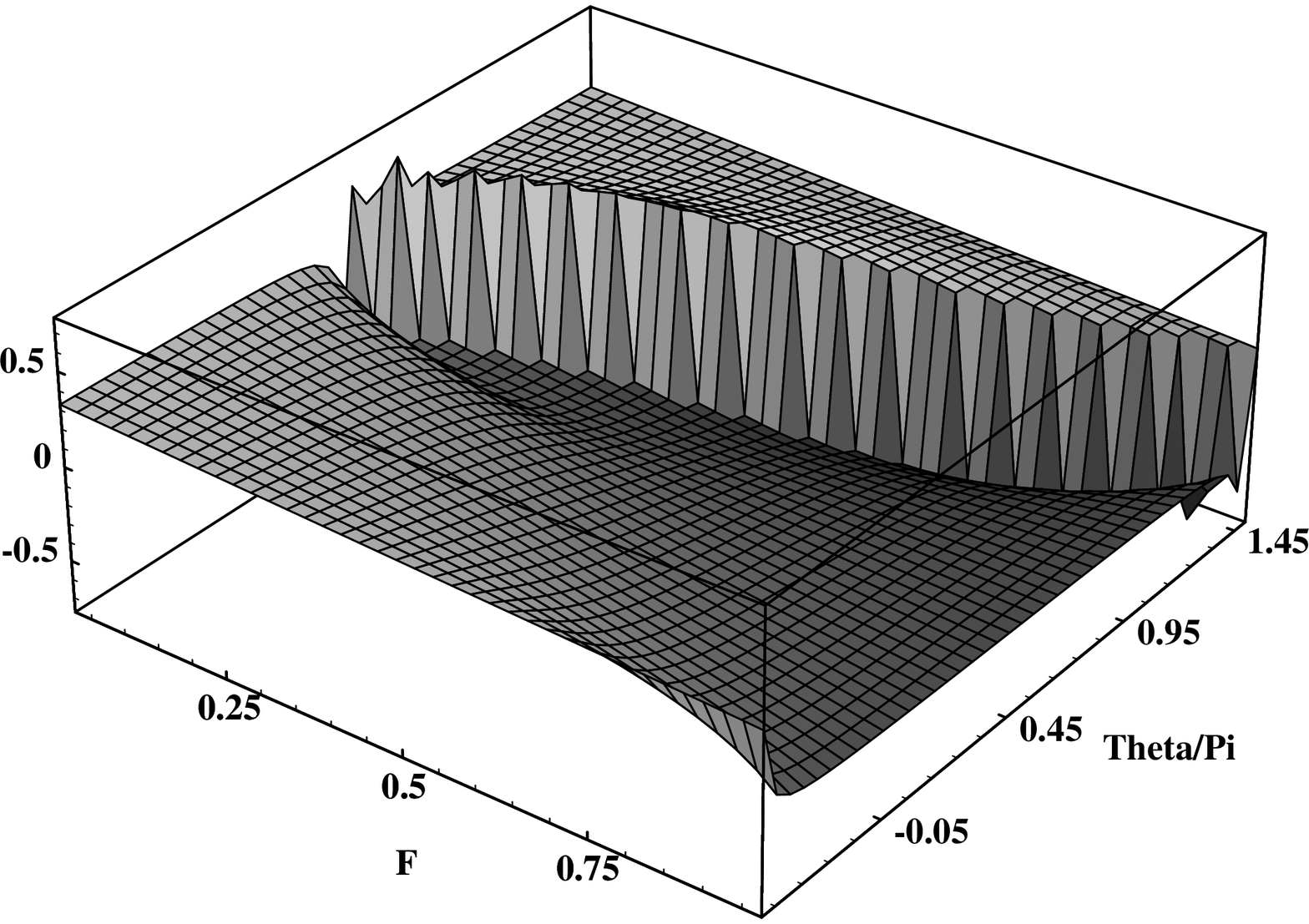}
\leavevmode
\end{center}
\vfill
\noindent Fig. 2.\qquad $2\pi|m|[e^2F(1-F)]^{-1}\Phi^{(I)}$ in the region
$-0.5<\Theta\pi^{-1}<1.5$,\quad $0<F<1$.
\newpage
\begin{center}
\thispagestyle{empty}
\epsfxsize=16cm
\epsffile{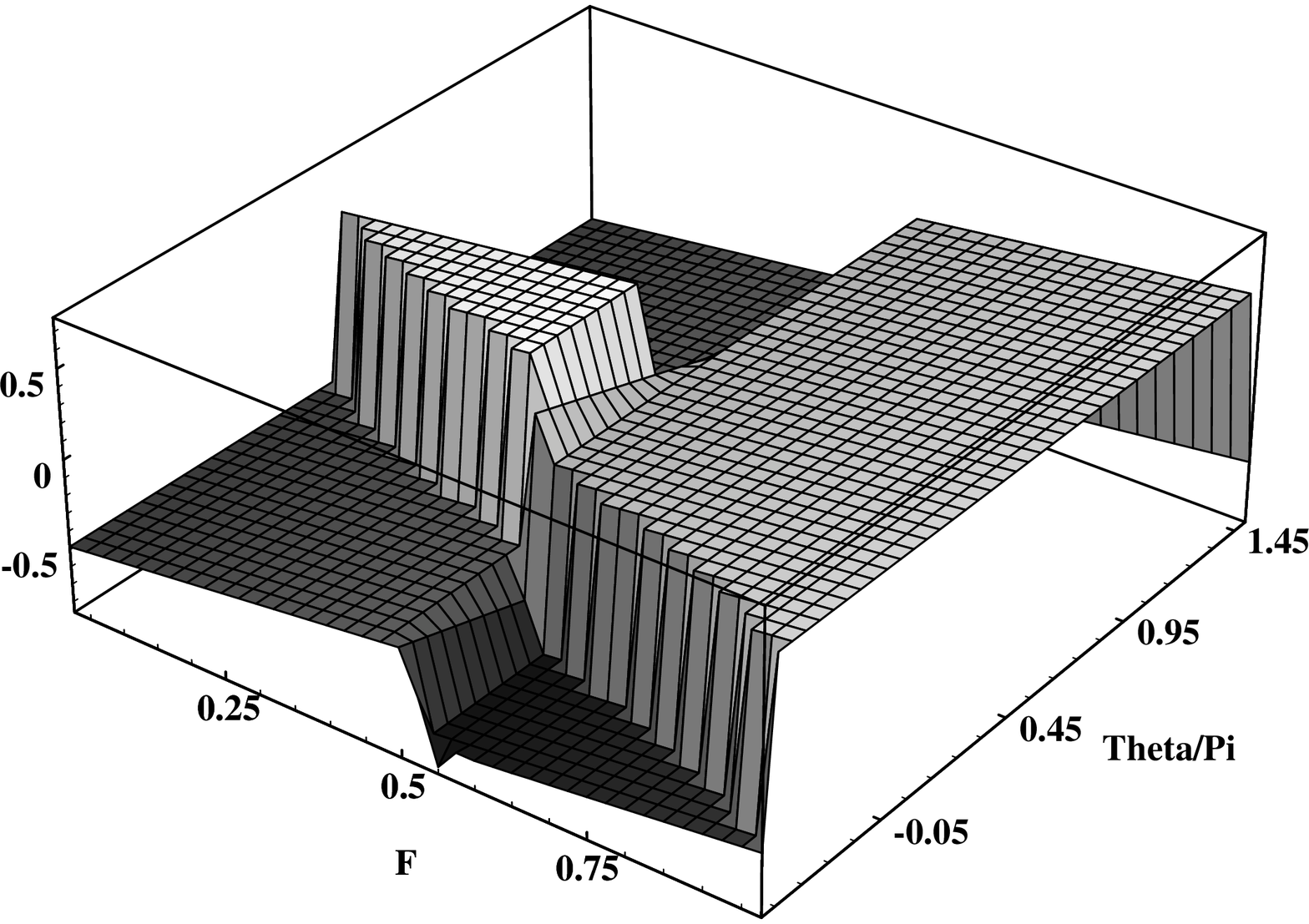}
\leavevmode
\end{center}
\vfill
\noindent Fig. 3.\qquad $Q^{(I)}$ in the region $-0.5<\Theta\pi^{-1}<1.5$, $0<F<1$
($m<0$).
\newpage
\begin{center}
\thispagestyle{empty}
\epsfxsize=16cm
\epsffile{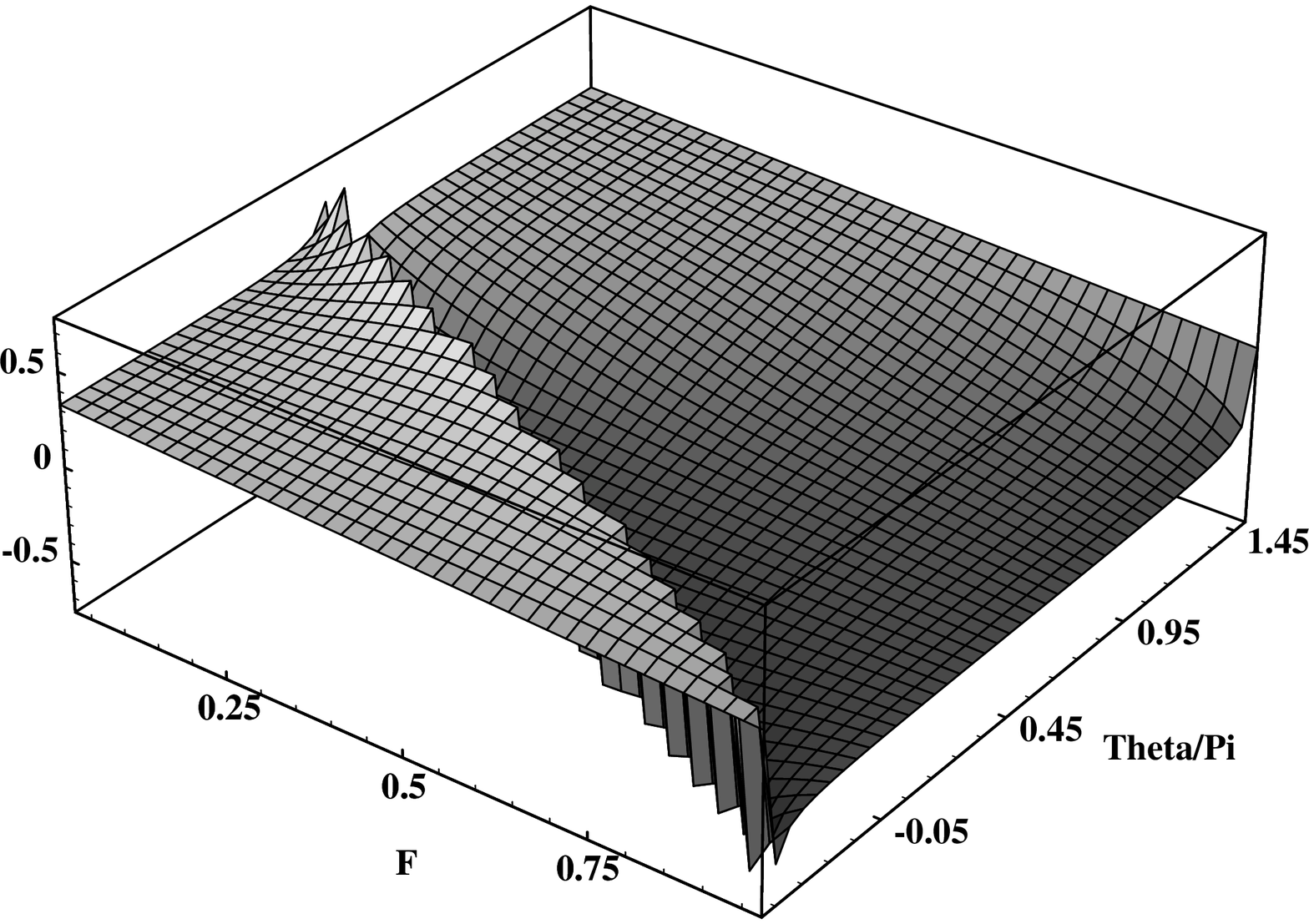}
\leavevmode
\end{center}
\vfill
\noindent Fig. 4.\qquad $2\pi|m|[e^2F(1-F)]^{-1}\Phi^{(I)}$ in the region
$-0.5<\Theta\pi^{-1}<1.5$,\quad $0<F<1$.

\end{document}